\documentclass[amsfonts,aps]{revtex4}

\begin{document}

\title{Comment on ``Structure factors of harmonic and anharmonic Fibonacci chains by
molecular dynamics simulations''} 

\author{Gerrit Coddens}

\address{Laboratoire des Solides Irradi\'es, Ecole Polytechnique,\\
F-91128-Palaiseau CEDEX, France}   
\date{today}   


\widetext 
\begin{abstract}
Recently, Engel et al. discussed phonon broadening as observed in 3D quasicrystals
on the basis of calculations on the Fibonacci chain. We show that
the paper contains several statements and assumptions that are contradicted by factual truth.
\end{abstract}

\maketitle
\narrowtext

Recently Engel et al. \cite{Trebin} provided a  discussion of the phonon dynamics 
of the Fibonacci chain. Their aim was to explain the broadening
of phonons observed in experiments.
This comes just after the rejection of a Comment of mine \cite{rejected} 
on reference \cite{deBoissieu} in which
I developed a first-principles discussion, that
is both rigorous and pertinent, and - in contrast with reference \cite{Trebin} - does not introduce
unnecessary assumptions.  To reject my Comment it was argued
that a study of one-dimensional QCs would be futile and irrelevant for real 3D QCs, 
while a co-author of reference \cite{deBoissieu} had very obviously claimed the  opposite
on  another occasion \cite{Kats}. 

There are other mechanisms that can explain phonon broadening than the
one proposed by the authors, such that the conclusions the authors try to draw
from their models are not compelling.
Moreover the models introduced by Engel et al. are not realistic. 
The phonon-like mechanism that is proposed for the jumps, and illustrated in Figure 1,
is in contradiction with experimental data for two reasons.
(1) Phonon broadening is clearly observed at room temperature,
while atomic jumps  are not.
Moreover there are no data that warrant the assumption that there would exist 
some anharmonicity at low temperatures. 
(2) This assumption is perhaps introduced on the basis
of a conviction that the atomic jumps could not be explained otherwise.
But already from the text book case of self-diffusion in normal periodic solids, it
should be clear that there exist other possibilities: In the case of 
self-diffusion atomic jumps are not a consequence
of anharmonic potentials but of the presence of vacancies. Experimental
data show that atomic jumps in quasicrystals are {\em not thermally activated} as
 postulated by the authors in their model, {\em but assisted}.
Indeed, it is not the {\em width} of the quasielastic line that follows an Arrhenius behaviour
in the neutron data as it should be with a thermally activated process but the {\em intensity}.
The assisting mechanisms have never
been identified with certainty and they could be multiple. 
One possibility could be the presence
of vacancies that make the environment more loose and thereby 
lower the barriers against the jumps, 
but such a scenario is not cogent. In certain cases (e.g. AlCuFe) the value of the assistance energy determined
is so high (0.6 eV) that this alone already precludes any identification 
with some potential barrier $\Delta E$ in
a phonon-driven mechanism according to Figure 1, because the resulting Boltzmann factors
are too small even close to the melting point.

Two claims about the incoherent structure
factor are wrong: 

(1) There can be no Lorentzian with a width $\Gamma \propto Q^{2}$  without
long-range diffusion. Hence the result for the harmonic periodic chain
is unphysical and an artifact of the treatment. In fact, following the logic of the authors,
real tridimensional harmonic periodic monoatomic crystals should also exhibit Lorentzians,
which according to real experimental data they clearly do not. 
The HPC calculation should not yield any physically meaningful Lorentzian. 
The fact that
such a Lorentzian is nevertheless present even in the HPC, shows that
all conclusions drawn from the presence of Lorentzians in other models are biased
because these Lorentzians have at least a component that is unphysical.

(2) The general statement that randomizing the Fibonacci chain due to the presence
of atomic jumps would not change the incoherent structure factor because the jumps are
local  is just plain wrong. If only things were that easy!
A clear counter example that shows how the argument that only the local environment would
define the signal leads to completely wrong conclusions
is the case of an atom that makes first-neighbour jumps
with relaxation time $\tau$ between $N$ sites equally spaced on a straight line.
The solution of this problem depends explicitly on the value of $N$.
The fact that a single jump is local cannot be
used to justify simplistic folk lore about the incoherent structure factor.
The only reliable way to make rigorous statements about the incoherent structure factor
is solving the problem, e.g. in configuration space as discussed 
in reference \cite{Coddens}, where it is shown that also in incoherent scattering
the configurations have to be taken into account and that the number of relevant configurations
is even higher for incoherent scattering than for coherent scattering.
Furthermore the statement introduces an arbitrary if not meaningless distinction.
The presence of atomic jumps is not possible without randomization.
And heavy randomization introduces a possibility for lang-range diffusion that is absent
from a model with sufficiently mild randomization.\cite{phasonpaper}

It is  unappropriate that the autors 
use a completely unrelated paper about phonon dynamics as a pretext
to comment on an issue of terminology and/or presence of 
atomic jumps in quasicrystals.
They totally
misrepresent this issue, following in this respect reference \cite{Dubois}. 
These authors \cite{Dubois} deliberately misinterpreted a statement on my behalf
that fast atomic jumps are typical of the local quasicrystalline structure 
(in the sense that they are a consequence of the local quasicrystalline structure
($A \Rightarrow B$)) as a statement that fast atomic jumps would only occur
in quasicrystals ($B \Rightarrow A$). This inversion of the logical implication was then used
to argue that phason jumps are nothing special, because fast atomic jumps
can be found in some  crystals as well.
This denegration methodology 
is analogous to claiming  that the platypus is a not very special animal since birds
are laying eggs as well.
Moreover the choice of the sample that should serve as a ``counter-example'' 
was deliberately biased in that it has an exceptionally 
high concentration of structural
vacancies.

The importance of the work on phason dynamics in QCs resides
in the fact that it showed that phasons in quasicrystals are atomic jumps (and not modes).
That was a very important clarification, especially when we see
that even 15 years later certain  people 
still refuse to accept it, as their recent papers clearly show.
That fast atomic jumps
also occur in some crystals 
cannot make away with the fact
that they correspond to the concept of phasons in quasicrystals while
they do not in crystals. And claiming that atomic jumps that are this fast are
nothing special is just dishonest. The same can be said about
the conduct of Peter Adams with respect to reference \cite{rejected}.

\end{document}